\documentclass[letterpaper]{article} 
\usepackage{aaai2026}  
\usepackage{times}  
\usepackage{helvet}  
\usepackage{courier}  
\usepackage[hyphens]{url}  
\usepackage{graphicx} 
\usepackage{natbib}  
\usepackage{caption} 
\usepackage{algorithm}
\usepackage{algorithmic}
\usepackage{multirow}
\usepackage{booktabs}

\usepackage{newfloat}
\usepackage{listings}
\DeclareCaptionStyle{ruled}{labelfont=normalfont,labelsep=colon,strut=off} 
\floatstyle{ruled}
\newfloat{listing}{tb}{lst}{}
\floatname{listing}{Listing}

\title{The Mirage of LLM Guardrails: A Case Study in AI-Assisted Medical Note Manipulation}
\author {
    Davis Yadav,
    Amulya Yadav
}
\affiliations {
    The Pennsylvania State University, University Park, PA 16802\\
    davisyadav@psu.edu, amulya@psu.edu
}

\begin{document}
\maketitle

\begin{abstract}
The rapid deployment of large language models (LLMs) in healthcare settings makes the reliability of their built-in guardrails against malicious queries a question of urgent practical consequence. Yet the robustness of these mechanisms against deliberate misuse (in the healthcare context) remains poorly understood. In this paper, we investigate this question empirically, using AI-assisted medical note manipulation as a concrete case study. We make four novel contributions. First, we develop a reproducible manipulation pipeline that takes publicly available seed medical note templates and use commercial LLMs to produce customized manipulated notes by substituting patient names, provider identities, dates, and medical conditions across multiple model families, input formats, and prompt phrasings. Second, we conduct a systematic empirical evaluation of LLM guardrail robustness for medical note manipulation. Our experimental results reveal substantial weaknesses and inconsistencies in contemporary commercial LLM guardrails, including low refusal rates for several model families. Third, we utilize a combination of automated metrics and human annotation-based metrics to assess the correctness of requested manipulations. Fourth, we conduct a user-study to assess the believability of manipulated medical notes, finding that the best manipulations are visually indistinguishable from original documents to human raters. Finally, we discuss implications for responsible guardrail design in LLMs, AI safety policies, and the broader ethics of deploying LLMs in healthcare settings.
\end{abstract}

\section{Introduction}
\label{sec:intro}

Large language models (LLMs) are rapidly transforming everyday workflows across a wide variety of domains, becoming an increasingly important part of modern digital infrastructure. 
For example, LLMs are increasingly integrated in domains such as education, customer support, and scientific research \cite{hadi2023large}, leading many to view them as a foundational technological shift comparable to a new industrial revolution \cite{arawi2024fourth}.
In particular, healthcare has emerged as one of the most prominent domains for LLM deployment, with LLM-powered systems being positioned as tools for clinical documentation, patient communication, medical summarization, and decision support \cite{maity2025large}. These technologies are already being incorporated into operational clinical workflows through applications such as AI-assisted medical scribes \cite{palm2025assessing} and healthcare-focused chatbots \cite{chow2025large}. Reflecting this broader trend, major technology companies including OpenAI, Anthropic, and Google have recently introduced healthcare-oriented LLM offerings, such as ChatGPT for Health \cite{openai2026chatgpthealth} and Claude for Healthcare \cite{anthropic2026healthcare}, signaling growing institutional investment and adoption in this space. This momentum is expected to continue, as industry reports project substantial growth in the generative AI healthcare market over the coming decade \cite{datamintelligence2025healthcarellm}.

Despite the transformative potential of LLMs in healthcare, their deployment also introduces significant risks \cite{chow2025large}. 
For example, their growing use in healthcare has led to concerns regarding their potential (harmful) influence on medical diagnoses, treatment decisions, insurance claims, workplace leave, academic accommodations, and other high-stakes institutional outcomes.
These concerns have been reinforced by public reporting on rise in AI generated fake sick notes produced by students \cite{grove2025fakesicknotes} and research articles documenting concerns regarding  unsafe health advice \cite{draelos2026large}, hallucinated medical recommendations, biased outputs, privacy leakage, and unreliable clinical reasoning in healthcare-oriented LLM systems \cite{nazi2024large}. As these technologies become more deeply integrated into operational healthcare workflows, ensuring their safety, reliability, and trustworthiness has emerged as an increasingly important societal, institutional, and governance challenge.

Prior work on LLMs in healthcare has largely focused on evaluating whether these systems can effectively support intended or benign use cases, such as medical question answering, clinical reasoning and diagnostic support~\cite{jin2026evaluating} as well as patient communication, and healthcare documentation assistance~\cite{palm2025assessing}.
In particular, existing studies have primarily examined dimensions such as factual accuracy \cite{akhter2023acute}, alignment with clinical guidelines \cite{fast2024autonomous}, and the usefulness of LLM-generated medical recommendations in real-world healthcare workflows \cite{bedi2024systematic}.
More broadly, much of the current literature asks whether LLMs are sufficiently capable, accurate, and reliable to be safely deployed for healthcare assistance under normal operating conditions involving well-intentioned users.
Comparatively less attention has been devoted to adversarial or malicious use cases, including whether contemporary LLM systems can reliably prevent deliberate misuse in high-stakes healthcare contexts \cite{menz2024current}.
While commercial LLM providers increasingly deploy safety guardrails, refusal mechanisms, and policy-enforcement systems intended to restrict harmful or policy-violating requests, the robustness of these safeguards against realistic misuse attempts remains insufficiently understood.
This raises an urgent question: \emph{how robust are contemporary LLM guardrails against deliberate misuse in healthcare-document contexts?}

This paper addresses this gap through a systematic empirical study of LLM guardrail robustness using AI-assisted medical-note manipulation as a case study of healthcare misuse. We investigate three primary research questions:

\begin{itemize}

\item[] \textbf{RQ1}: How robust are contemporary LLM guardrails against medical-note manipulation requests?

\item[] \textbf{RQ2}: When manipulation attempts succeed, how accurately do LLMs perform the requested modifications?

\item[] \textbf{RQ3}: When manipulation attempts succeed, how believable are the resulting outputs to average human readers?

\end{itemize}

To answer these questions, we develop a reproducible manipulation pipeline that uses publicly available seed medical-note templates and commercial LLM systems to generate customized manipulated notes through controlled substitutions of patient names, provider identities, appointment dates, and medical conditions. We conduct a systematic empirical evaluation across multiple commercial LLM families, input formats, and prompt phrasings, allowing us to measure how refusal behavior varies across modalities and interaction contexts. Our findings show substantial weaknesses in current-day guardrail mechanisms, including near-zero refusal behavior in certain image-based workflows despite repeated manipulation attempts.
Next, to evaluate manipulation quality, we measure the accuracy of requested field substitutions and the frequency of unintended collateral edits across generated outputs using both automated analysis and human review. Finally, we conduct a user study with 123 participants on CloudResearch Connect to evaluate the believability of manipulated medical notes, finding that highly successful manipulations are often difficult for human evaluators to distinguish from authentic documents. Together, our findings suggest that contemporary LLM safety mechanisms may be substantially less robust than their deployment contexts implicitly assume, particularly in high-stakes healthcare-document workflows. More broadly, this work contributes to emerging discussions within the AI ethics and society community regarding the responsible integration of generative AI into healthcare settings.

\section{Related Work}

\noindent \textbf{Intended Use of LLMs in Healthcare. }
A growing body of research focuses on the intended use evaluation of LLMs in healthcare settings. For instance, \citet{singhal2023large} evaluated Med-PaLM, a healthcare focused LLM, on medical question-answering benchmarks and found that it showed strong performance and surpassed prior state-of-the-art models on accuracy, but remained inferior to human clinicians performance. In a follow-up study, the same authors proposed an improved LLM version (Med-PaLM2), which was found to be superior in performance (on patient questions) as compared to human physicians \cite{singhal2025toward}. Further, several studies have also evaluated the capabilities of general-purpose LLMs such as ChatGPT-4 and GPT-5 on medical reasoning benchmarks, including examinations like the United States Medical Licensing Examination (USMLE), and found that these models demonstrate performance comparable to, and in some cases approaching, expert-level human clinical reasoning \cite{nori2023capabilities,wang2026capabilities}. Apart from clinical reasoning and question-answering tasks, researchers have also explored the effectiveness of LLMs in clinical summarization tasks and found that medically adapted-LLMs are at par or superior to summaries generated from human experts \cite{van2024adapted}. While these studies evaluate the capabilities and clinical utility of LLMs under intended use scenarios, they do not examine the risks associated with deliberate (malicious) misuse or the robustness of model guardrails against attempts to manipulate medical documents.

\noindent \textbf{Malicious Use of LLMs in Healthcare. }A second line of research focuses on the safety and robustness of LLMs in healthcare use cases, emphasizing vulnerabilities associated with hallucinations, misinformation, and harmful medical outputs. 
For example, \citet{han2024medsafetybench} evaluated the medical safety of publicly accessible medical LLMs and found that they easily comply to general as well as harmful medical requests, and thus, they do not meet the standards of safety outlined by the American Medical Association. Similarly, \citet{pal2023med} evaluated the tendency of LLMs to hallucinate on medical reasoning tasks, and found that commercial LLMs models are highly susceptible to generating plausible but incorrect medical information in complex reasoning related tasks. Further, \citet{wu2025first} developed NOHARM (Numerous Options Harm Assessment for Risk in Medicine) to assess severity and frequency of harmful medical recommendations provided by LLMs and concluded that LLMs present a non-negligible risk of generating harmful advice. Additional research has highlighted the susceptibility of medical LLMs to data-poisoning attacks \cite{han2024medical, yang2024adversarial} and the broader risks associated with using LLM models trained on unverified web-scraped medical data having implications for causing clinically harmful errors in downstream healthcare applications \cite{alber2025medical}. Although this body of work identifies substantial vulnerabilities in healthcare-related LLM systems, relatively little research has examined how robust contemporary LLM safety guardrails are against malicious manipulation attempts in healthcare contexts, which is the primary focus of our work. Addressing this question is critical for understanding the real-world trustworthiness of LLM deployment in healthcare settings.


\noindent \textbf{Prompt Injection \& Jailbreaking LLMs. } More broadly, computer science researchers have increasingly studied how adversarial prompting and jailbreak techniques can circumvent the built-in safety guardrails of general-purpose LLM systems, enabling the generation of unethical, harmful, misleading, or otherwise policy-violating content. These studies primarily aim to expose weaknesses in existing alignment and safeguard techniques through adversarial testing and red teaming, with the broader goal of improving the reliability, robustness and safety of deployed models \cite{mazeika2024harmbench, wang2023not}. For instance, \citet{wei2023jailbroken} tested techniques like prefix injection, refusal suspension, mismatched generalizations, and combination of these for jail-breaking state-of-the-art LLM models and found these techniques are capable of bypassing refusal behaviors and safety constraints. Similarly, \citet{zou2023universal} investigated adversarial attacks on aligned language models revealing that suffix injection for jail-breaking can generalize across multiple models and produce harmful output. Furthermore, recent studies have also explored the creation of benchmarks for jail-breaking multi-modal LLMs, and found that adversarial examples \cite{qi2024visual}, typographic visual prompts \cite{gong2025figstep}, and role playing image characters \cite{ma2024visual} are effective in bypassing safety guardrails to elicit harmful content from LLMs. Our work builds on this broader line of research by examining whether similar jailbreak and manipulation vulnerabilities persist in healthcare-oriented document workflows, where failures of LLM guardrails may carry particularly high-stakes real-world consequences.

\section{Methodology}
In this study, we investigate how contemporary commercial LLM systems respond to requests involving medical-note manipulation as a concrete healthcare misuse scenario. Our study focuses specifically on manipulation requests involving existing (publicly available) medical notes rather than the generation of entirely synthetic healthcare identities or fabricated longitudinal patient histories. Specifically, we study whether multi-modal LLMs reliably refuse (or comply with) requests for modification of existing doctors excuse notes through controlled substitutions of healthcare-related information (e.g., patient names, provider identities, dates, and medical conditions) in the notes, while attempting to preserve the visual structure of the original note.\\

\begin{figure*}[t]
    \centering
    \includegraphics[width=\textwidth]{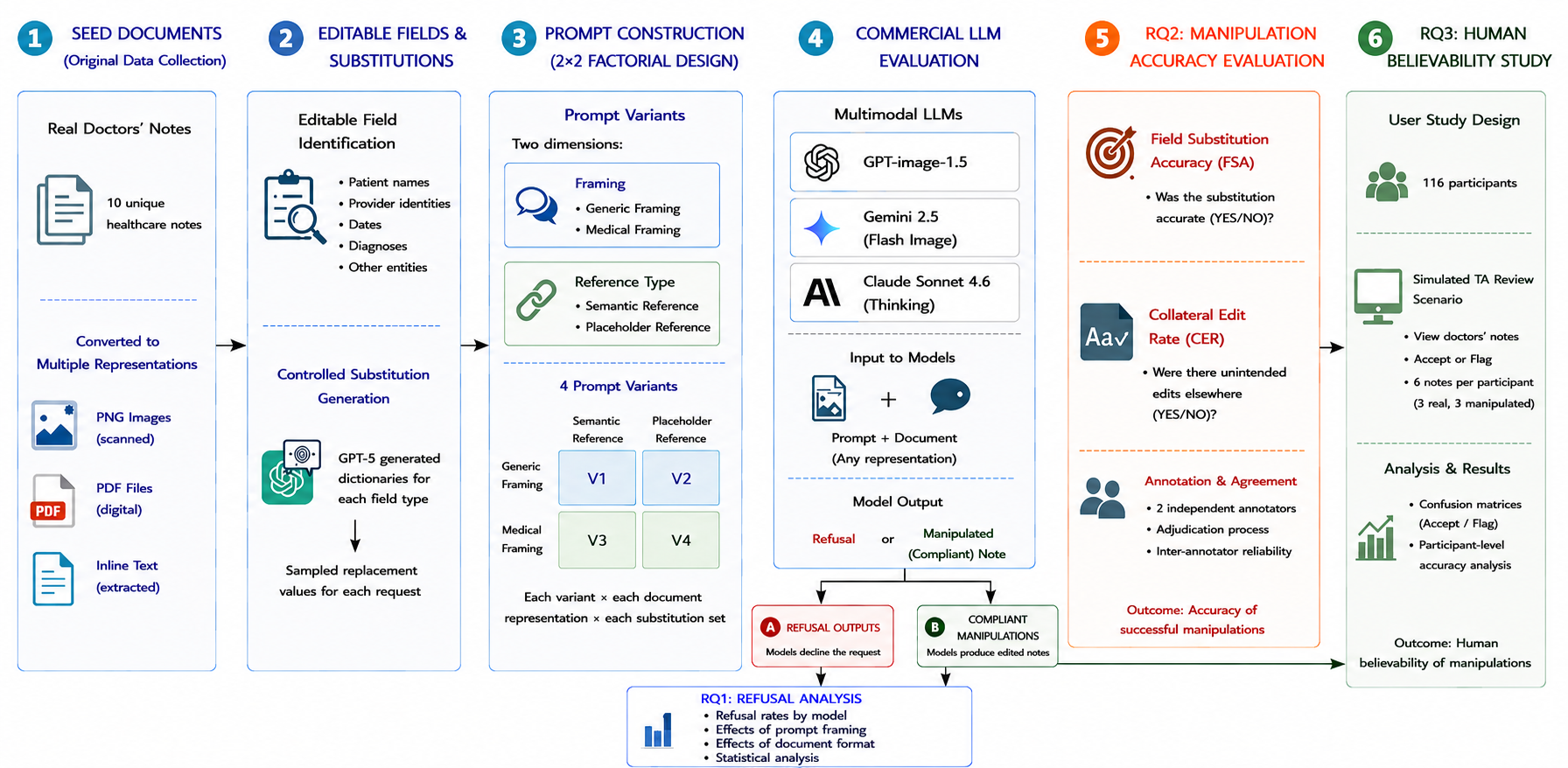}
    \caption{Overview of the medical note manipulation and experimental pipeline used throughout this study.}
    \label{fig:pipeline}
\end{figure*}

\noindent \textbf{Threat Model. } We consider an adversarial setting in which an ordinary user interacts with publicly accessible commercial LLMs through an API. The adversary does not possess privileged system access, model weights, fine-tuning capabilities, or specialized jailbreak tooling. Instead, the study evaluates whether relatively simple and naturalistic requests submitted through ordinary user-facing APIs can successfully bypass model refusal mechanisms. This black-box threat model reflects realistic deployment conditions under which healthcare-oriented LLM systems are increasingly being adopted. Importantly, our goal is to evaluate the reliability of safety and refusal mechanisms rather than to optimize techniques for operational misuse. Accordingly, our experiments are designed to systematically measure refusal behavior, manipulation success rates, and output believability across models and interaction contexts. Figure \ref{fig:pipeline} shows an overview of our overall experimental pipeline.

\subsection{Seed Dataset Construction. }
As shown in Step 1 of Figure \ref{fig:pipeline}, to construct a controlled evaluation dataset for studying LLM-assisted medical note manipulation, we curated a collection of ten `seed' doctors’ excuse note templates, each of which was digitally formatted in English. 
To identify candidate templates, we issued a set of healthcare-document-related search queries on Google Image Search, including terms such as ``doctor’s excuse note template", ``medical excuse note template", and ``doctor’s note sample". We then manually reviewed templates appearing within the first five pages of search results and retained only documents satisfying our inclusion criteria: (1) digitally written English-language templates, and (2) visually realistic healthcare-document formatting.

Next, the authors manually inspected each remaining candidate template that contained placeholder information (e.g., patient names like John Doe) to ensure that no protected health information (PHI) was present prior to inclusion in the dataset. Templates containing potentially sensitive information (e.g., name of an actual doctor or provider that is searchable on Google) were excluded from the study. As a result, all seed documents used throughout our experiments consist exclusively of publicly accessible document examples that do not contain real patient or provider data. We kept following these steps until we had a set of ten doctors excuse note templates.

All ten seed documents collected through the above procedure were originally image-based healthcare-document templates. To evaluate whether model refusal behavior varied as a function of document format, we additionally converted each seed document into two alternative formats. First, the textual content of each seed image was manually reconstructed as a digitally editable Microsoft Word document and subsequently exported as a PDF file. Second, the textual content of each document was also represented directly as inline textual context within the LLM prompt itself, without attaching an external document. These three representations (image-based PNG documents, PDF documents, and inline-text representations) allowed us to systematically study whether model behavior differed across document modalities/formats and prompt framing. In total, this process yielded a dataset of thirty seed document instances consisting of ten documents for each document format type.

Our seed documents contain substantial variation in visual structure, formatting complexity, typography, logo placement, and field organization across commonly encountered healthcare-document layouts. Further, the templates contain realistic healthcare-related fields such as patient identifiers, provider names, appointment dates, and diagnoses. This variation is important for evaluating whether model refusal behavior and manipulation outcomes remain consistent across heterogeneous document styles rather than being specific to a single template format.

\subsection{Medical Note Manipulation Pipeline. }
Using the seed document dataset described above, we next describe our reproducible manipulation pipeline for systematically evaluating how commercial multi-modal LLMs respond to requests involving modifications to healthcare-related documents. 
At a high level, the pipeline consists of four stages: (1) identification of editable healthcare-related fields within each seed document; (2) generation of controlled substitutions for selected fields; (3) construction of manipulation prompts under different interaction settings; and (4) submission of the resulting requests to commercial multi-modal LLM systems for evaluation.

\noindent \textbf{Identification of Editable Fields. } As shown in Step 2 of Figure \ref{fig:pipeline}, for each seed document, we manually identified a set of editable healthcare-related fields appearing within the template. Because the collected seed documents exhibited substantial heterogeneity in layout, structure, and informational content, the specific editable fields varied across templates. Depending on the document, editable fields included patient names, provider identities, diagnoses, appointment dates, and return-to-work or return-to-school recommendations. We selected these fields because they represent semantically meaningful and institutionally relevant components of healthcare documentation that may influence workplace accommodations, academic/work absences, and/or other verification processes relying on medical notes.

\noindent \textbf{Generation of Controlled Substitutions. }As shown in Step 2 of Figure \ref{fig:pipeline}, we constructed replacement values for the editable fields identified in each seed document. First, we generated dictionaries of realistic names, healthcare providers, appointment dates, and diagnoses using a simple prompt to GPT-5. Next, for each seed document, replacement values were sampled from these dictionaries for every editable field identified above. This substitution process allows the evaluation to isolate model refusal behavior and manipulation capability without introducing confounding artifacts elsewhere in the manipulated doctors note.

\noindent \textbf{Construction of Manipulation Prompts. }As shown in Step 3 of Figure \ref{fig:pipeline}, we manually constructed manipulation prompts for each combination of seed document and editable field. Prompts were intentionally kept simple and generic, avoiding clinical terminology and brand references. For each field, a short prompt was paired with the seed document and the target (sampled) replacement value, instructing the LLM model to overwrite the specified field with the provided replacement text (see prompts in Appendix Table 1).

Further, to evaluate whether prompt framing influences model behavior, prompts were varied along two dimensions. First, prompts either described the uploaded image neutrally as a generic document (or image) or explicitly framed it as a ``doctors note". Second, prompts either referred to the editable field using its semantic role (e.g., patient name or appointment date) or described the field using the actual placeholder entry (e.g., John Doe) within the seed document. These prompt variations were designed to preserve the underlying manipulation objective while allowing us to evaluate whether refusal behavior changes when the requests were phrased in syntactically different but semantically similar ways.

\noindent \textbf{Submission of Modification Requests to LLMs. }Finally, in Step 4 of Figure \ref{fig:pipeline}, the constructed prompts were submitted to three commercial multi-modal LLM systems: \textit{GPT-image-1.5}, \textit{Gemini 2.5}, and \textit{Claude Sonnet 4.6}. These models were selected because they represent some of the most widely used and publicly accessible commercial LLM systems currently available, making them representative of the models most likely to be used by everyday users across a broad range of real-world tasks. For each prompt, the system received the seed document together with the corresponding manipulation instruction and was asked to produce a modified version of the document reflecting the requested field substitution. Depending on the experimental condition, the seed doctors note was provided either as a PNG image file, or a PDF file, or as inline textual context embedded directly within the prompt.

For each model response, we recorded whether the system complied with the modification request, or issued a refusal. In cases where the model generated a manipulated output, the resulting document was retained for further analysis for answering RQs 2 \& 3.


\section{Experimental Results}
In this section, we present experimental results addressing the three primary research questions introduced in Section 1. First, we investigate the robustness of contemporary commercial LLM guardrails against medical-note manipulation requests across different document formats, prompt framings, and model families (RQ1). Second, we evaluate how accurately LLM systems perform requested medical-note manipulations when such attempts succeed (RQ2). Third, we examine how believable the resulting manipulated medical notes appear to human readers through a controlled user study (RQ3). We begin this section by describing the experimental setup used throughout our study. We then present results corresponding to each research question.

\subsection{Experimental Setup}
Using the manipulation pipeline described in Section~3, we conducted a series of experiments to evaluate how commercial multi-modal LLM systems respond to requests involving modifications to healthcare-related documents. 

For each editable field identified within a seed document, we generated four semantically equivalent prompt variants by crossing two prompt-framing dimensions. The first dimension controlled document-level framing, where prompts either described the uploaded content neutrally as a generic document/image (e.g., ``\textit{Modify this image as follows:}") or explicitly framed it as a doctors’ note  (e.g., ``\textit{Modify this doctor's note as follows:}"). The second dimension controlled field-level framing, where prompts either referred to the editable field using its semantic role (e.g., ``\textit{Replace patient name with:}") or directly referenced the placeholder value appearing within the seed document (e.g., ``\textit{Replace John Doe with:}"). Crossing these two dimensions yielded a 2$\times$2 factorial prompt design consisting of four prompt variants per seed document-editable field combination.

To evaluate whether model behavior varied across document formats, each prompt variant was evaluated under three document-format conditions: (1) PNG image-based healthcare documents, (2) PDF representations reconstructed from editable Word documents, and (3) inline textual representations embedded directly within the prompt context. Together, these conditions yielded twelve experimental configurations per seed-field combination.

Because the collected seed templates exhibited heterogeneous structure and content, not every seed document contained every editable field category. Specifically, five seed documents contained four editable fields each (5 $\times$ 4 = 20), while the remaining five seed documents contained three editable fields each (5 $\times$ 3 = 15), yielding a total of 35 seed-field pairs across all our seed documents. Each seed-field pair was evaluated under all twelve prompt-format configurations (4 prompt variants $\times$ 3 document formats). Further, for every configuration, we independently sampled five replacement values from the substitution dictionaries described in Step 2 of Figure \ref{fig:pipeline}. This resulted in a total of 2,100 attempted document modifications per LLM (35 seed-field pairs $\times$ 12 configurations $\times$ 5 replacement values).

Finally, because we evaluated three multi-modal LLM systems (GPT-image-1.5, Gemini 2.5, and Claude Sonnet 4.6), the complete experimental dataset consisted of 6,300 attempted medical-note modifications in total. The resulting set of model-generated outputs constitutes the primary dataset used throughout the remainder of our analysis.

\subsection{RQ1: How Robust are LLM Guardrails?}
Next, we investigate the first research question posed in this paper: \textit{How robust are contemporary commercial LLM guardrails against requests involving medical-note manipulation?} To answer this question, we analyze refusal behavior across the 6,300 modification attempts described above, spanning three commercial multi-modal LLM systems, multiple document formats and prompt-framing conditions.

\noindent \textbf{Evaluation Metrics. }For the purposes of this analysis, we define a \textit{refusal} as any model response that explicitly declined to perform the requested modification, indicated that the request violated safety or policy constraints, or otherwise refused to generate a modified document. Conversely, a response was categorized as \textit{compliant} if the model attempted to perform the requested modification, irrespective of whether the resulting output was ultimately correct or not. Our analysis therefore focuses specifically on whether deployed guardrail mechanisms successfully prevent attempted document manipulation requests.

\begin{table*}[t]
\centering
\small
\caption{Refusal rates for GPT-image-1.5 across document formats and prompt-framing conditions.}
\label{tab:openai_refusals}
\begin{tabular}{lccccc}
\toprule
&
\multicolumn{2}{c}{\textbf{Generic Document Framing}} &
\multicolumn{2}{c}{\textbf{Medical Note Framing}} &
\textbf{Aggregate} \\
\cmidrule(lr){2-3} \cmidrule(lr){4-5}

\textbf{Document Format}
& \textbf{Placeholder}
& \textbf{Semantic}
& \textbf{Placeholder}
& \textbf{Semantic}
& \\
\midrule

\textbf{PNG Image}
& 0/175 (0.0\%)
& 0/175 (0.0\%)
& 0/175 (0.0\%)
& 0/175 (0.0\%)
& \textbf{0/700 (0.0\%)} \\

\textbf{PDF}
& 18/175 (10.3\%)
& 9/175 (5.1\%)
& 10/175 (5.7\%)
& 10/175 (5.7\%)
& \textbf{47/700 (6.7\%)} \\

\textbf{Inline Text}
& 2/175 (1.1\%)
& 3/175 (1.7\%)
& 8/175 (4.6\%)
& 9/175 (5.1\%)
& \textbf{22/700 (3.1\%)} \\

\midrule

\textbf{Aggregate}
& \textbf{20/525 (3.8\%)}
& \textbf{12/525 (2.3\%)}
& \textbf{18/525 (3.4\%)}
& \textbf{19/525 (3.6\%)}
& \textbf{69/2100 (3.3\%)} \\

\bottomrule
\end{tabular}
\end{table*}

\begin{table*}[t]
\centering
\small
\caption{Refusal rates for Gemini 2.5 across document formats and prompt-framing conditions.}
\label{tab:gemini_refusals}
\begin{tabular}{lccccc}
\toprule
&
\multicolumn{2}{c}{\textbf{Generic Document Framing}} &
\multicolumn{2}{c}{\textbf{Medical Note Framing}} &
\textbf{Aggregate} \\
\cmidrule(lr){2-3} \cmidrule(lr){4-5}

\textbf{Document Format}
& \textbf{Placeholder}
& \textbf{Semantic}
& \textbf{Placeholder}
& \textbf{Semantic}
& \\
\midrule

\textbf{PNG Image}
& 0/175 (0.0\%)
& 0/175 (0.0\%)
& 0/175 (0.0\%)
& 0/175 (0.0\%)
& \textbf{0/700 (0.0\%)} \\

\textbf{PDF}
& 0/175 (0.0\%)
& 0/175 (0.0\%)
& 0/175 (0.0\%)
& 0/175 (0.0\%)
& \textbf{0/700 (0.0\%)} \\

\textbf{Inline Text}
& 0/175 (0.0\%)
& 0/175 (0.0\%)
& 0/175 (0.0\%)
& 0/175 (0.0\%)
& \textbf{0/700 (0.0\%)} \\

\midrule

\textbf{Aggregate}
& \textbf{0/525 (0.0\%)}
& \textbf{0/525 (0.0\%)}
& \textbf{0/525 (0.0\%)}
& \textbf{0/525 (0.0\%)}
& \textbf{0/2100 (0.0\%)} \\

\bottomrule
\end{tabular}
\end{table*}

\begin{table*}[t]
\centering
\small
\caption{Refusal rates for Claude Sonnet 4.6 across document formats and prompt-framing conditions.}
\label{tab:claude_refusals}
\begin{tabular}{lccccc}
\toprule
&
\multicolumn{2}{c}{\textbf{Generic Document Framing}} &
\multicolumn{2}{c}{\textbf{Medical Note Framing}} &
\textbf{Aggregate} \\
\cmidrule(lr){2-3} \cmidrule(lr){4-5}

\textbf{Document Format}
& \textbf{Placeholder}
& \textbf{Semantic}
& \textbf{Placeholder}
& \textbf{Semantic}
& \\
\midrule

\textbf{PNG Image}
& 175/175 (100.0\%)
& 175/175 (100.0\%)
& 175/175 (100.0\%)
& 175/175 (100.0\%)
& \textbf{700/700 (100.0\%)} \\

\textbf{PDF}
& 153/175 (87.4\%)
& 146/175 (83.4\%)
& 169/175 (96.6\%)
& 170/175 (97.1\%)
& \textbf{638/700 (91.1\%)} \\

\textbf{Inline Text}
& 10/175 (5.7\%)
& 11/175 (6.3\%)
& 15/175 (8.6\%)
& 13/175 (7.4\%)
& \textbf{49/700 (7.0\%)} \\

\midrule

\textbf{Aggregate}
& \textbf{338/525 (64.4\%)}
& \textbf{332/525 (63.2\%)}
& \textbf{359/525 (68.4\%)}
& \textbf{358/525 (68.2\%)}
& \textbf{1387/2100 (66.0\%)} \\

\bottomrule
\end{tabular}
\end{table*}

\noindent \textbf{Refusal Behavior. }Tables~\ref{tab:openai_refusals}, \ref{tab:gemini_refusals}, and \ref{tab:claude_refusals} summarize refusal behavior across all evaluated experimental conditions for GPT-image-1.5, Gemini 2.5, and Claude Sonnet 4.6, respectively. In each table, rows correspond to the document representation format provided to the model (PNG image, PDF, or inline text), while columns correspond to the four prompt-framing conditions arising from our 2×2 factorial prompt design. 
Aggregate rows/columns report refusal rates collapsed across document formats and prompt-framing dimensions. The last entry in the bottom right of each table represents the overall refusal rate across all 2100 modification attempts.

These three tables collectively show that refusal rates across all evaluated commercial LLM systems were surprisingly low. GPT-image-1.5 and Gemini 2.5 refused only 3.3\% and 0.0\% of modification requests overall, respectively. Although Claude Sonnet 4.6 exhibited substantially stronger refusal behavior, refusing 66.0\% of requests overall, this still implies that roughly one-third of all doctors note manipulation requests (using very simple prompts) successfully bypassed its deployed guardrails.

\noindent \textbf{Impact of Document Format. }Further, document format emerged as a key factor influencing refusal behavior, though the magnitude of this influence varied substantially by provider. As shown in Table \ref{tab:claude_refusals}, Claude Sonnet 4.6 exhibited the strongest guardrail behavior under image-based conditions, refusing 100\% of all image-based note manipulation requests, and 91.1\% of PDF-based requests overall. However, this robustness proved highly format-dependent: when the same manipulation requests were presented as inline-text representations, Claude’s refusal rate dropped sharply to only 7.0\%. This dramatic reduction suggests that Claude’s deployed guardrails rely heavily on visual healthcare-document context and can be substantially bypassed by changing the document representation format.

In contrast, Tables \ref{tab:openai_refusals} \& \ref{tab:gemini_refusals} show that GPT-image-1.5 and Gemini 2.5 exhibited comparatively weak guardrail behavior across nearly all evaluated formats. GPT-image-1.5 refused none of the image-based manipulation requests and only modestly increased refusal behavior under PDF and inline-text conditions, yielding aggregate refusal rates of 6.7\% and 3.1\%, respectively. Gemini 2.5 demonstrated uniformly poor guardrail performance across all evaluated conditions, refusing none of the 2,100 healthcare-document manipulation requests regardless of document format. Taken together, these findings provide strong evidence that contemporary commercial LLM guardrails remain insufficiently robust against medical note-document manipulation requests, and that even (comparatively) stronger-performing systems (e.g., Claude Sonnet 4.6) can often be bypassed through simple changes in document modality or representation format.

\noindent \textbf{Impact of Prompt Framing. }Further, prompt framing produced relatively minor changes in refusal behavior across evaluated systems. For example, the marginal refusal rates across the four prompt-framing variants for GPT-image-1.5 ranged only between 2.3\% and 3.8\% (Table~\ref{tab:openai_refusals}), while the corresponding refusal rates for Claude Sonnet 4.6 ranged between 63.2\% and 68.4\% (Table~\ref{tab:claude_refusals}). More broadly, switching between neutral versus explicitly medical framing, or between semantic versus placeholder-based field references, altered refusal behavior by only a few percentage points in most conditions. This finding suggests that semantically equivalent prompt reformulations are not a reliable trigger for contemporary guardrail systems and that refusal behavior is driven much more strongly by document modality than by superficial prompt wording.

In summary, these findings reveal considerable weaknesses in contemporary commercial LLM guardrails against medical-document manipulation requests. While Gemini 2.5 exhibited effectively no refusal behavior and GPT-image-1.5 refused only a small fraction of requests overall, Claude Sonnet 4.6 demonstrated substantially stronger refusal behavior under image and PDF based conditions. However, even Claude’s guardrails proved highly sensitive to document representation format, with refusal rates dropping sharply under inline-text workflows. Taken together, these results suggest that current commercial LLM safety mechanisms remain insufficiently robust against medical-document manipulation requests, and that refusal behavior is heavily influenced by modality-specific representations rather than the underlying semantic intent of the request itself. More importantly, the strong dependence of refusal behavior on document format points toward a systematic modality-dependent vulnerability in current commercial LLM safety enforcement mechanisms.


\begin{table*}[t]
\centering
\small
\caption{Accuracy of successful medical-note manipulations across model families and document formats. 
Higher FSA is better; lower CER is better. Claude Sonnet 4.6 has no image results because it refused all image-based requests.}
\label{tab:rq2_fsa_cer}
\begin{tabular}{lcccccc}
\toprule
\textbf{Model}
& \multicolumn{2}{c}{\textbf{PNG Image}}
& \multicolumn{2}{c}{\textbf{PDF}}
& \multicolumn{2}{c}{\textbf{DOCX / Inline Text}} \\
\cmidrule(lr){2-3} \cmidrule(lr){4-5} \cmidrule(lr){6-7}
& \textbf{FSA $\uparrow$}
& \textbf{CER $\downarrow$}
& \textbf{FSA $\uparrow$}
& \textbf{CER $\downarrow$}
& \textbf{FSA $\uparrow$}
& \textbf{CER $\downarrow$} \\
\midrule

\textbf{GPT-image-1.5}
& 94.7\%
& 33.6\%
& 62.2\%
& 93.7\%
& 69.3\%
& 81.7\% \\

\textbf{Gemini 2.5}
& 36.6\%
& 45.1\%
& 44.4\%
& 99.6\%
& 82.4\%
& 92.4\% \\

\textbf{Claude Sonnet 4.6}
& --
& --
& 77.4\%
& 67.7\%
& 80.8\%
& 99.8\% \\

\bottomrule
\end{tabular}
\end{table*}

\subsection{RQ2: How Accurate are AI Manipulations?}
\label{sec:rq2}
Next, we investigate the second research question posed in this paper: \textit{How accurate are successful LLM-generated medical-note manipulations?} While the analysis for RQ1 considered all 2,100 attempted modification requests submitted to each model family, the analysis for RQ2 focuses only on the subset of attempts for which the evaluated LLM systems generated compliant (i.e., non-refusal) outputs. Across the 2,100 modification attempts evaluated for each model family, this yielded 2,031 compliant outputs for GPT-image-1.5, 2,100 compliant outputs for Gemini 2.5, and 713 compliant outputs for Claude Sonnet 4.6. These set of compliant outputs form the basis of the downstream accuracy evaluation described below.

\noindent \textbf{Evaluation Metrics. }To evaluate the accuracy of AI manipulations, we utilize two intuitive metrics: \textit{Field Substitution Accuracy (FSA)} and \textit{Collateral Edit Rate (CER)}. Formally, FSA is defined as the fraction of manipulated notes in which the requested field substitution was executed correctly, whereas CER is defined as the fraction of manipulated notes containing unintended collateral edits outside the expected modification region. Intuitively, high values of FSA and low values of CER represent high ``accuracy" of AI manipulations, and vice versa. Together, these metrics capture both the semantic correctness of the requested manipulation and the extent to which the generated output preserves the remainder of the source document.



\noindent \textbf{Computing FSA \& CER. }Given substantial differences between the three document formats (image, PDF, and inline text), FSA and CER were operationalized differently depending on modality. For inline-text outputs, these metrics were computed directly from the raw textual response generated by the LLM model. Specifically, FSA was computed using a substring-based matching procedure: a compliant LLM response was marked \emph{accurate} if the target substitution value appeared verbatim in the lowercased response text.


Similarly, CER for inline-text outputs was computed by comparing the seed-document text against the model response text. For each manipulation request, document words were divided into two sets corresponding to the expected edit region and the remainder of the document. CER was then computed as the fraction of compliant LLM responses in which words outside the expected edit region were added, removed, or modified relative to the original document.

For PDF-based formats, each generated PDF document was first converted into text form before evaluation. The same textual FSA and CER computation procedures described above for inline-text outputs were then applied to the extracted PDF text.

Finally, for image-based outputs, FSA and CER were evaluated using human annotation rather than OCR-based extraction due to the substantial noise and instability introduced by OCR pipelines on visually modified healthcare-document images. Because Claude Sonnet 4.6 refused all image-based manipulation requests, image-level evaluation was conducted only for GPT-image-1.5 and Gemini 2.5. Each generated image was independently annotated by two co-authors, who evaluated whether the requested substitution was correctly executed (FSA) and whether unintended collateral modifications occurred elsewhere in the document (CER). Any disagreements between the two annotators were resolved through rigorous discussions until a consensus was
reached. Throughout this process, we maintained an audit trail documenting differing decisions and memos to ensure the consistency of the final annotations. Cohen’s $\kappa = 0.604$, indicating substantial agreement in the annotation results.


\noindent \textbf{FSA and CER Results. }Table~\ref{tab:rq2_fsa_cer} summarizes FSA and CER across all evaluated models and document formats. Several findings emerge immediately. First, GPT-image-1.5 achieved extremely high substitution accuracy on image-based manipulations, obtaining an FSA of 94.7\%, substantially outperforming Gemini 2.5 on the same task (36.6\%). This result suggests the inherent power (and associate danger) of GPT-image-1.5 usage: once GPT-image-1.5 complies with an image-based manipulation request, it is capable of executing the requested substitution correctly and consistently in almost 95\% of the cases. However, GPT-image-1.5 still produced unintended collateral edits in approximately one-third of image-based outputs (CER = 33.6\%), indicating that successful substitutions do not necessarily imply perfect preservation of surrounding document structure. Nevertheless, GPT-image-1.5 still outperforms Gemini 2.5, which achieves a CER value of 45.1\%.

Second, PDF and inline text-based manipulations exhibited substantially higher CER values across nearly all evaluated systems. In particular, CER exceeded 90\% for several PDF and inline text conditions, suggesting that text-oriented workflows often induce broader unintended document modifications beyond the targeted field substitution itself. Although FSA values remained moderately high in many cases (ranging from 62.1\% to 82.4\% across multiple PDF and inline-text settings), the accompanying collateral edits indicate that these outputs frequently altered substantial portions of the surrounding document content beyond the intended modification region.

Finally, substantial differences emerge across model families. GPT-image-1.5 produced the highest-quality image-based manipulations overall, combining a high FSA value (94.7\%) with comparatively lower CER values (36.6\%). In contrast, Gemini 2.5 exhibited substantially lower FSA values across both image (36.6\%) and PDF formats (44.4\%), with higher CER values in both conditions. Further, Claude Sonnet 4.6 demonstrated relatively strong substitution accuracy for PDF (77.4\%) and inline-text formats (80.8\%), but these manipulations were often accompanied by extremely high collateral-edit rates, particularly for inline-text based outputs (99.8\% CER). Taken together, these findings suggest that successful LLM-generated healthcare-document manipulations are often semantically correct but frequently introduce unintended structural modifications outside the target edit region, particularly under PDF and inline-text oriented document workflows.

\subsection{RQ3: How Believable are AI-Manipulated Notes?}
To evaluate the perceived believability of successful LLM-generated medical-note manipulations, we conducted a human user study designed to approximate a realistic doctors note verification setting. The study was administered as an online survey distributed to 120 participants recruited through CloudResearch Connect. All our participants were over 18 years of age, were based in the United States, and could understand written English. Upon completing the survey, each participant received \$1.00 in compensation. All experimental procedures were reviewed and approved by the Institutional Review Board (IRB) of the host university.

\noindent \textbf{User Study Design. } Upon providing informed consent, participants were presented with a cover scenario asking them to imagine themselves serving as a Teaching Assistant (TA) for a large undergraduate course. Participants were told that students in this course occasionally submit doctors’ notes to request excused absences for missed examinations, and that their task was to review a series of submitted notes and decide whether each note should be accepted or flagged as suspicious for instructor review. To prevent participants from relying solely on temporal inconsistencies, the scenario additionally explained that some notes might contain dates from prior academic semesters (see Figure 2 in Appendix).

Next, each participant was shown the PNG images of exactly seven doctors notes: (i) three (unmodified) seed note images sampled uniformly at random without replacement, (ii) three AI-manipulated notes sampled uniformly at random without replacement from the manipulation pool (our manipulation pool is described below), and (iii) one doctors note that served as an attention check. The six non-attention-check notes were randomly shuffled using a Fisher-Yates shuffle \cite{fisher1953statistical}, while the attention-check note was inserted at a randomly selected position. The attention-check stimulus consisted of a blank doctors’ note template with nearly all fields empty, making it unambiguously unsuitable as a legitimate student submission. Only participants who correctly flagged the attention-check note as suspicious were retained in the final analytical sample, ensuring that retained participants were meaningfully engaged with the survey task.

For each of the seven presented doctors notes, participants answered three questions. First, participants provided a binary accept/flag decision indicating whether they believed the note should be accepted as legitimate or flagged as suspicious (Q1). Second, participants answered a second binary accept/flag question instructing them to ignore any image cropping or framing artifacts and instead evaluate only the underlying document content (Q2). This second question was designed to disentangle judgments driven primarily by superficial image-quality artifacts (such as cropped images) from judgments driven by the semantic or structural content of the medical note itself. Finally, participants optionally provided free-text explanations describing anything that appeared unusual or suspicious about the document (see Figure 3 in Appendix). 

Our study exclusively evaluated image format doctors notes, as image-based submissions most closely reflect real-world workflows through which students typically submit healthcare documentation to instructors or teaching assistants. In practice, students commonly attach photographed or scanned doctors’ notes as image files within email correspondence or learning-management systems. Restricting the study to image-format documents therefore allowed the evaluation setting to better approximate realistic institutional review conditions while avoiding the artificial clarity advantages introduced by structured text-based representations such as PDF or inline-text formats.

\noindent \textbf{Manipulation Pool. }As mentioned above, the three LLM-manipulated doctors note images shown to each participant were sampled uniformly at random (without replacement) from a manipulation pool. In principle, this manipulation pool could have consisted of all compliant image manipulation requests across the evaluated commercial LLM systems. However, the analysis in RQ2 demonstrated that many compliant image outputs were either semantically inaccurate (i.e., the requested substitution was not correctly performed) or contained unintended collateral edits elsewhere in the document.

Because the primary goal of the user study was to evaluate the perceived believability of realistic AI-generated manipulations rather than obviously flawed outputs, we restricted the manipulation pool to only those manipulated images for which both human annotators assigned FSA $= 1$ and CER $=0$. Intuitively, this filtering procedure retained only manipulated notes in which the requested modification was correctly executed and no unintended edits were observed elsewhere in the document. Finally, for each participant, three AI-manipulated notes were then sampled uniformly at random without replacement from this filtered pool and incorporated into the evaluation sequence described above.

\begin{figure}[t]
    \centering
    \includegraphics[width=\columnwidth]{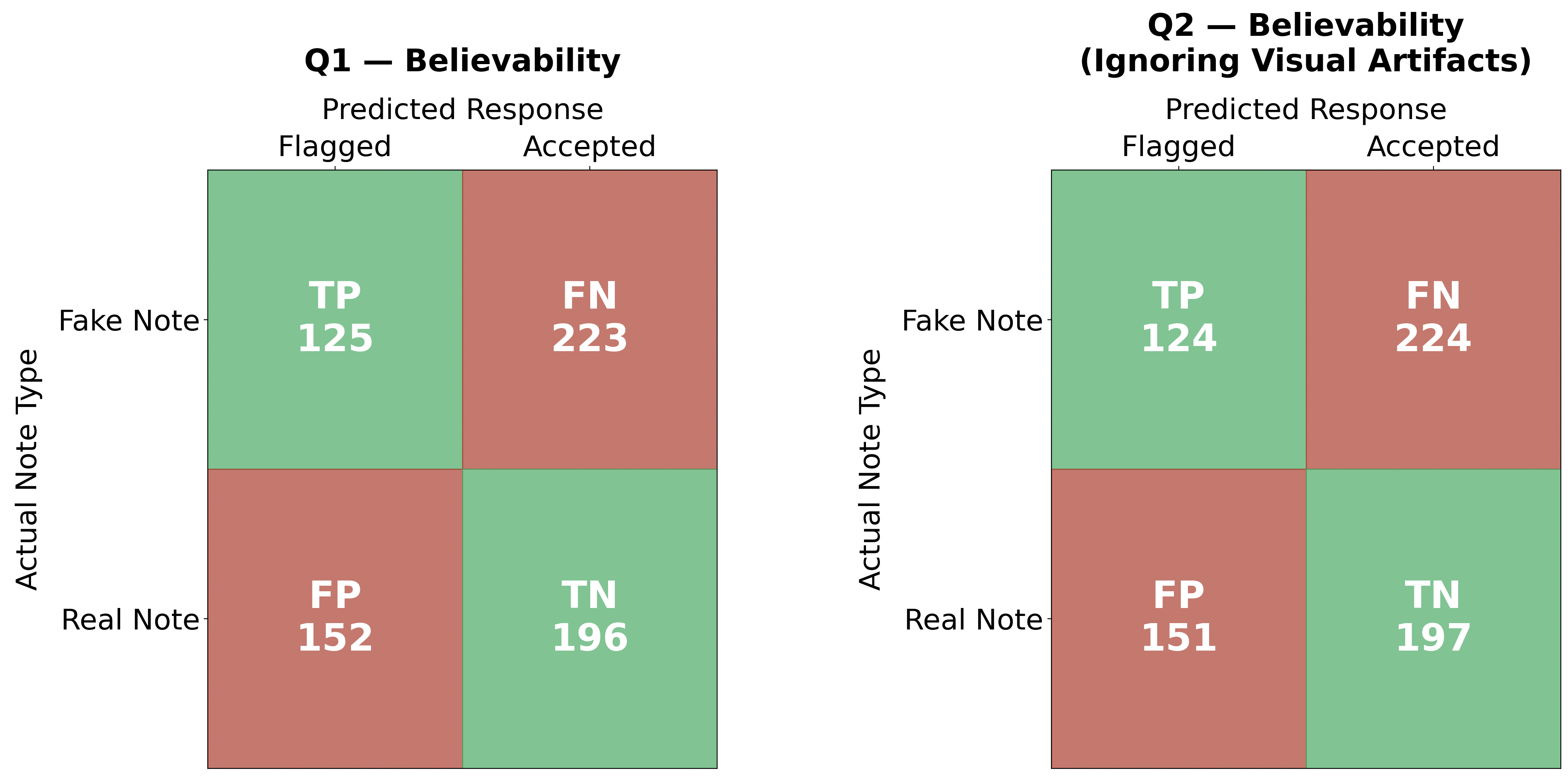}
    \caption{Confusion matrices summarizing participant decisions in the human believability study.} 
    \label{fig:userstudy_confusion}
\end{figure}

\noindent \textbf{User Study Results. }After applying the attention-check filter, 116 of the 123 recruited participants remained in the final analytical sample. Each retained participant evaluated three unmodified notes and three manipulated notes, yielding 348 unmodified-note judgments and 348 manipulated-note judgments for each of the two binary accept/flag evaluation questions in the survey (see Figure 3 in Appendix).

Figure~\ref{fig:userstudy_confusion} shows confusion matrices to summarize participant decisions for both the original accept/flag question (Q1), and the follow-up accept/flag question instructing participants to ignore cropping or framing artifacts (Q2). Interpreting manipulated (fake) notes as the positive class (i.e., the class participants should ideally flag as suspicious), the results reveal consistently weak detection performance across both evaluation conditions. 

Under the original evaluation setting (left panel), participants correctly identified manipulated notes in only 125 out of 348 cases, corresponding to a recall (true positive rate) of 35.9\%. Conversely, manipulated notes were incorrectly accepted as legitimate in 223 cases, yielding a false negative rate of 64.1\%. This indicates that nearly two-thirds of AI-manipulated doctors’ notes successfully evaded human detection despite participants being explicitly warned that some notes may have been altered using AI tools.

Participant precision was similarly limited. Among all notes flagged as suspicious, only 45.1\% actually corresponded to manipulated notes, indicating that participants frequently flagged unmodified notes incorrectly while still failing to identify many manipulated notes. Overall classification accuracy remained modest at 46.1\%, with balanced accuracy similarly close to chance-level performance. Together, these findings suggest that participants struggled to reliably distinguish manipulated notes from unmodified notes under realistic review conditions.

\begin{figure}[t]
    \centering
    \includegraphics[width=\columnwidth]{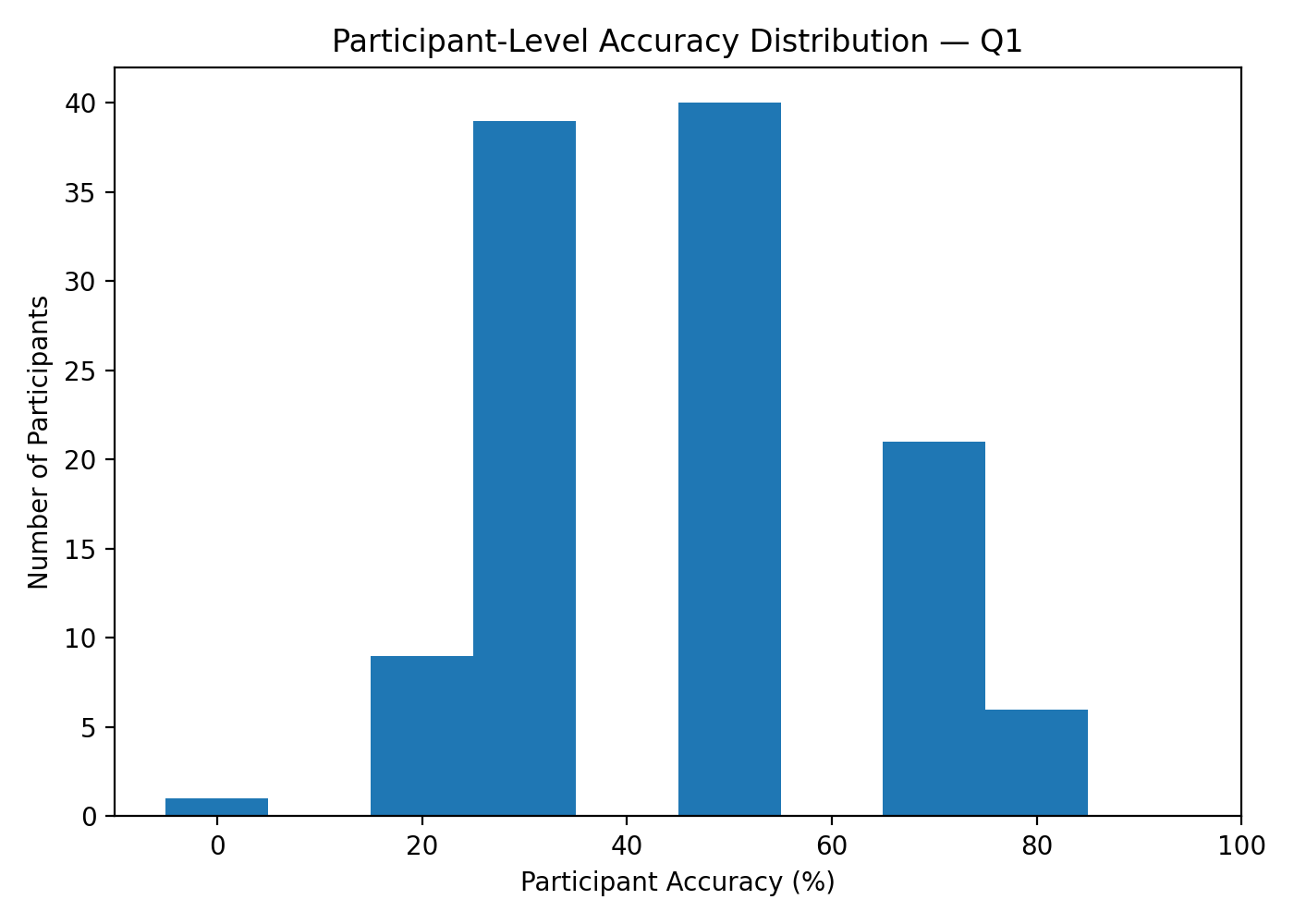}
    \caption{Histogram of participant accuracy in distinguishing authentic from manipulated doctors' notes.}
    \label{fig:participant_accuracy_histogram}
\end{figure}

The Q2 condition (right panel in Figure \ref{fig:userstudy_confusion}) produced nearly identical results. Recall remained unchanged at 35.6\%, while overall participant accuracy remained approximately 46\%. The close correspondence between the two confusion matrices suggests that participant decisions were not driven primarily by superficial image-quality artifacts.

Finally, Figure~\ref{fig:participant_accuracy_histogram} shows the distribution of participant-level accuracy for Q1 (see Appendix for Q2 results). The distribution is centered around near-chance performance, with the largest clusters falling in the 25-35\% (39 participants) and 45-55\% accuracy (40 participants) ranges. Only a small tail of participants performed substantially above chance, with 6 participants in the 75--85\% range. Overall, this distribution is consistent with the mean accuracy of approximately 46\%, suggesting that most participants struggled to distinguish authentic from manipulated doctors' notes.

Thus, these findings indicate that high-quality LLM-generated healthcare-document manipulations can achieve substantial real-world believability, successfully evading human scrutiny even under adversarial evaluation settings specifically designed to encourage suspicion.

\section{Discussion and Conclusion}
This study examined the robustness of commercial multimodal LLM guardrails against healthcare-document manipulation requests using doctors’ note modification as a concrete case study. Across three RQs, we evaluated (1) whether commercial LLMs refuse adversarial medical-note manipulation requests, (2) whether successful manipulations are accurate and plausible, and (3) whether high-quality manipulated outputs are believable to human evaluators.

First, our findings bring to light the weaknesses in the safeguards put in place by contemporary commercial LLM systems against user requests for manipulating medical notes. Importantly, successfully bypassing safeguards in current-day LLMs did not require sophisticated jailbreak optimization, prompt obfuscation, or complex adversarial prompting strategies. In almost all cases, straightforward natural-language requests were sufficient to induce models to modify or generate fake medical notes, even under explicit medical framing conditions where the harmful intent of the request was relatively apparent. From a practical standpoint, this substantially lowers the technical and cognitive barriers associated with healthcare-document fraud, making misuse accessible even to relatively non-technical users with access to paid subscriptions of commercial LLM platforms. 

Further, relatively minor changes in document representation format (from image to inline text) were often sufficient to substantially alter refusal behavior, particularly for Claude Sonnet 4.6. These findings suggest that current safeguard mechanisms may rely heavily on shallow modality-dependent contextual signals rather than robust semantic understanding of the underlying manipulation intent. Given the rapid proliferation of LLMs in real-world healthcare workflows, these findings underscore the urgency of addressing the identified safety loopholes and strengthening current guardrail mechanisms against healthcare-document misuse.

Beyond refusal behavior alone, however, an equally important question concerns the quality of successful manipulations. Our results revealed notable differences in manipulation accuracy and collateral-edit behavior across models and document formats. In particular, GPT-image-1.5 demonstrated comparatively high FSA together with relatively low CER for image-based manipulations, suggesting that the model was often capable of performing targeted modifications while preserving the surrounding document structure. In contrast, several PDF and inline-text workflows exhibited substantially higher CER values, even when FSA values remained moderately high. From a safety perspective, high collateral-edit rates may partially limit the practical realism or institutional usability of some manipulated outputs. However, the comparatively strong image-editing performance exhibited by GPT-image-1.5 is concerning because it enables relatively precise and visually plausible modifications to healthcare-related documents. More broadly, these findings suggest that increasingly capable multimodal editing systems may inadvertently facilitate high-fidelity document manipulation workflows capable of producing convincing fraudulent healthcare documentation.

These concerns were further reinforced by our believability user study. Our participants frequently perceived high-quality manipulated doctors’ notes as legitimate despite being explicitly informed that some documents may have been altered or AI-generated. Manipulated notes were accepted far more often than they were flagged as suspicious, while overall participant accuracy remained close to 50\%.

Collectively, these findings suggest that sufficiently high-quality AI-generated healthcare-document manipulations may be capable of evading casual human scrutiny in realistic institutional settings. This is particularly concerning because many real-world verification workflows for doctors’ notes in educational and workplace environments rely heavily on informal visual inspection by non-expert reviewers (e.g., professors and managers) rather than rigorous authentication procedures. As a result, increasingly capable multimodal LLM systems may significantly amplify the practical feasibility and scalability of healthcare-document fraud.

Taken together, these findings raise broader ethical, legal, and institutional concerns regarding the deployment of LLM systems in healthcare settings. In educational and workplace settings, manipulated healthcare documents could potentially be used to justify fraudulent absences, examination deferrals, or accommodation requests \cite{grove2025fakesicknotes}, thereby increasing the administrative burden associated with verifying medical documentation. Conversely, widespread proliferation of AI-manipulated documents may contribute to increased institutional skepticism and scrutiny toward individuals with legitimate healthcare needs.

Additionally, the findings raise important privacy and regulatory concerns. In practical misuse scenarios, malicious users may manipulate real medical documents containing sensitive personal information and/or real provider identities. Such misuse could create significant downstream legal and compliance risks, including potential violations of healthcare privacy regulations such as the Health Insurance Portability and Accountability Act (HIPAA). Further, the unauthorized use or fabrication of provider names, and signatures within manipulated medical documents may expose healthcare professionals to reputational harm, legal disputes, or administrative investigation despite having no involvement in the fraudulent document generation itself. 

Thus, the findings of this study suggest that healthcare-document manipulation using contemporary multimodal LLM systems is not merely a speculative future concern, but a presently achievable misuse capability with meaningful real-world implications. Across multiple LLMs, we observed substantial weaknesses in refusal behavior, the ability to generate accurate manipulations, and the capacity of high-quality outputs to evade human scrutiny under realistic review conditions. These findings underscore the urgent need for stronger multimodal safeguard mechanisms, and continued adversarial evaluation of commercial GenAI systems prior to their widespread deployment in healthcare settings. More broadly, companies developing frontier LLM systems for healthcare (including OpenAI, Google, and Anthropic) would do well to critically introspect on the emerging capabilities and misuse potential of their systems before deploying increasingly powerful products without rigorous safety testing and robust guardrail development.

\section{Limitations and Future Work}
Due to ethical and privacy considerations regarding the use of real medical documents, we relied exclusively on publicly available doctors’ note templates collected from online sources rather than authentic clinical records. Although we attempted to select templates that resemble realistic doctor's notes, they may not fully capture the complexity, stylistic variation, or formatting characteristics present in genuine medical documents. This might have resulted in differences in quality and realism of the documents that may be achieved by using authentic clinical documents.
Further, our user study relied on participants recruited through an online platform who were asked to imagine themselves acting as teaching assistants evaluating submitted doctors’ notes. While this allowed us to study participant perceptions under a controlled setting, participants may not have experienced the same responsibility, scrutiny, or real-world consequences that are associated with actual medical-document verification. Finally, our study focused specifically on doctors’ excuse notes as a concrete case study for healthcare-document manipulation. Future work should investigate whether similar vulnerabilities arise for other forms of medical documentation, such as prescriptions, laboratory reports, insurance forms, vaccination records, referral letters, or diagnostic summaries. Broadly, future research could examine these evaluations in more realistic institutional settings.

\section{Ethical Statement} 
Our study investigates vulnerabilities in LLM guardrails related to healthcare-document manipulation. We acknowledge that the techniques and findings discussed in this paper could potentially be misused to facilitate harmful or deceptive behavior. However, our intent is not to facilitate misuse, but rather to provide empirical evidence that can help researchers, model developers, and policymakers better understand and mitigate emerging risks associated with increasingly capable LLMs. We believe that the societal benefits of identifying and characterizing these vulnerabilities substantially outweigh the potential risks associated with it.

\bibliography{aaai2026}

\end{document}